\documentclass[12pt]{article}
\usepackage{amssymb}
\usepackage{latexsym}
\usepackage{amsmath}
\usepackage{graphicx}

\usepackage{esdiff}
\usepackage{mathtools}

\def\vp{\varphi}
\def\al{\alpha}
\def\nb{\nabla}
\def\th{\theta}
\def\e{\eta}
\def\k{\kappa}
\def\l{\lambda}
\def\L{\Lambda}
\title{\LARGE{Slowly rotating black hole solution in the scalar-tensor theory with nonminimal derivative coupling and its thermodynamics}}
\author{M. M. Stetsko\footnote{E-mail: mstetsko@gmail.com}\
\\
  {\small Department for Theoretical Physics, Ivan Franko National University of Lviv,}\\
{\small 12 Drahomanov Str., Lviv, UA-79005, Ukraine
         }}

\begin{document}
\maketitle

\abstract{We obtain a slowly rotating black hole solution in the scalar-tensor theory of gravity with nonminimal derivative coupling to the Einstein tensor. Properties of the obtained solution have been examined carefully. We also investigate the thermodynamics of the given black hole. To obtain thermodynamic functions, namely its entropy we use the Wald procedure which is suitable for quite general diffeomorphism-invariant theories. The applied approach allowed us to obtain the expression for entropy and the first law of black hole thermodynamics. Having introduced thermodynamic pressure which is related to the cosmological constant we have examined thermodynamics of the black hole in the so called extended phase space. The extended phase space and specifically chosen scalar ``charge'' allowed us not only to obtain the generalized first law but also derive the Smarr relation. The behaviour of black hole's temperature, heat capacity and Gibbs free energy shows a lot of similarities with the behaviour of the corresponding values for Schwarzschild-AdS black hole in standard General Relativity.}

\section{Introduction}
Einstein\rq{}s theory of General Relativity is extremely successful theory whose predictions are in tight agreement with huge amount of  precision  experiments, especially for the weak-field or slow-motion regime \cite{Will_LRR2014,Berti_CQG2015}. At the same time some  strong field predictions still are difficult to be verified but recently made detection of gravitational waves \cite{Abbott_PRL16,Abbott_PRL16_2,Abbott_PRL17} opens new way for examination of strong-field gravity. Black holes are among the ideal objects to check the predictions of General Relativity. It should be noted that several important issues such as origin of curvature singularities, cosmological constant problem, dark energy/dark matter issue, higher order curvature corrections bring the idea of modification of General Relativity \cite{Clifton_PhysRept2012}. Among number of possible generalizations of General Relativity so called Horndeski gravity \cite{Horndeski_IJTP74} has been attracting a lot of interest for recent years.  Proposed few decades ago, Horndeski gravity \cite{Horndeski_IJTP74} represents the most general scalar tensor theory of gravity where field equations of motion are of the second order over derivatives. Horndeski gravity was rediscovered in the context of Galileon theories, namely the scalar-tensor theories with Galilean symmetry in flat space-time. The equivalence of Galileon theories in a space of arbitrary dimension \cite{Deffayet_PRD09} to Horndeski gravity in four dimensions was established \cite{Kobayashi_PTEP11}. It was  shown that Horndeski gravity can be obtained from higher dimensional Lovelock theory with help of Kaluza-Klein procedure \cite{Acoleyen_PRD11,Charmousis_LNP15}. 
It should also be noted that tensor-multiscalar theories \cite{Damour_CQG92,Horbatsch_CQG15} and multiscalar versions of Horndeski gravity \cite{Deffayet_PRD10,Padilla_JHEP13,Charmousis_JHEP14,Ohashi_JHEP15} can be formulated. 

Despite its appealing features and numerous applications Horndeski gravity in its general setup remains a bit cumbersome and it is important to maintain some of the attractive features in particular cases of the general Horndeski theory. One of the most interesting particular cases is the so called  theory with nonminimal derivative coupling where the scalar field inherited from the Horndeski theory is nonminimally coupled to gravity. We also note that  nonminimal derivatively coupled terms of the scalar field and gravity appear in low energy effective action of string theory \cite{Metsaev_NPB87,Meissner_PLB97,Cartier_PRD01} and ghost-free nonlinear massive gravity \cite{deRham_PRD11}. The theory with nonminimal derivative coupling was also applied to vast area of problems in cosmology \cite{Almendola_PLB93,Capozziello_AnnPh00}. In particular, inflationary cosmological solutions were considered in \cite{Almendola_PLB93,Capozziello_AnnPh00}. Exact cosmological solutions with derivative coupling were also examined \cite{Sushkov_PRD09}.  Quintessence and phantom cosmology solutions were considered and different types of evolutionary scenarios were obtained \cite{Saridakis_PRD10}. Dynamics of scalar field in a cosmological model with nonminimal derivative coupling was investigated in \cite{Granda_JCAP10,Gao_JCAP10,Sadjadi_PRD11}. Role of nonminimal coupling in the accelerated  expansion at late times was examined in \cite{Granda_JCAP10_1}. Slow-roll inflation was considered \cite{Germani_PRL10,Germani_PRL11}. Cosmological models with nonminimal coupling and additional power-law potential and their stability were investigated \cite{Skugoreva_PRD13}. Reheating process during rapid oscillations of inflation in the framework of nonminimally derivatively coupled theory was considered in \cite{Sadjadi_JCAP13,Dalianis_JCAP17}.  Curvaton model was also examined \cite{Feng_PLB14,Feng_PRD14,Qiu_EPJC17}.

Investigation of black holes and other compact object is also very interesting issue. For the first time a black hole in the theory with nonminimal derivative coupling without cosmological constant in four dimensional case was considered in \cite{Rinaldi_PRD12}  where a static solution was obtained and examined. Cosmological constant was taken into account in the paper \cite{Minamitsuji_PRD2014} where four and five dimensional solutions were obtained and thermodynamics of corresponding black holes was investigated. Stealth Schwarzschild and partially self-tuned dS-Schwarzschild solutions were examined in the work \cite{Babichev_JHEP14}. Four and multidimensional solution was also investigated in \cite{Anabalon_PRD14,Cisterna_PRD14}.  Black hole solution in a more general shift-symmetric Horndeski theory was examined \cite{Kobayashi_PTEP14}. BTZ-type black hole solution and its thermodynamics were considered in \cite{Bravo-Gaete_PRD14}. Thermodynamics of uncharged and charged black holes in case of nonminimal derivate coupling was investigated in \cite{Feng_JHEP15,Feng_PRD15}. The existence of black hole hair in Horndeski theory was examined  in \cite{Sotiriou_PRL14}. Four dimensional slowly rotating black hole in Horndeski theory was studied \cite{Maselli_PRD2015}. Warped three dimensional AdS black hole solution was obtained \cite{Giribet_PRD15}. Stable black hole solution in shift-symmetric Horndeski theory was examined in \cite{Tretyakova_CQG17}. Neutron and boson stars in the theory with nonminimal derivative coupling were investigated \cite{Cisterna_PRD15,Cisterna_PRD16,Brihaye_PRD16,Verbin_PRD18}. Black holes with nonminimal coupling and Gauss-Bonnet term were examined \cite{Antoniou_PRL18,Antoniou_PRD18}.

In our work we consider multidimensional slowly rotating black hole  solution in the theory with nonminimal derivative coupling. This work is organized in the following way. In the second section we write the equations of motion for the theory with nonminimal derivative coupling taking into account the assumption about slow rotation. In the third section we solve the written equations of and investigate obtained solution. In the forth section we obtain black hole temperature and entropy and examine its thermodynamics. Finally, the fifth section contains some conclusions. 

\section{Field equations for the system with nonminimal derivative coupling}
We start from a relation for action which consists of two parts, namely standard Einstein-Hilbert term plus cosmological constant and the second part which includes terms with minimal and  nonminimal derivative coupling with some  scalar field. The resulting action can be written in the form:
\begin{equation}\label{action}
S=\int d^{n+1}x\sqrt{-g}\left( R-2\Lambda-\frac{1}{2}\left(\alpha g^{\mu\nu}-\eta G^{\mu\nu}\right)\partial_{\mu}\vp\partial_{\nu}\vp \right)
\end{equation} 
where $g_{\mu\nu}$ denotes the the metric, $g=det(g_{\mu\nu})$ is the determinant of the mentioned metric, $R$ and $G_{\mu\nu}$  are the Ricci scalar and the Einstein tensor for the metric $g_{\mu\nu}$ correspondingly, $\Lambda$ is the cosmological constant, $\vp$ denotes the scalar field  coupled to gravity and finally $\al$ and $\e$ are minimal and nonminimal coupling constants respectively.  

Varying the action (\ref{action}) and having used the principle of the least action one can derive equations of motion. The equations for the gravitational field  take the following form: 
\begin{equation}\label{eins_eq}
G_{\mu\nu}+\Lambda g_{\mu\nu}=\frac{1}{2}(\alpha T^{(1)}_{\mu\nu}+\eta T^{(2)}_{\mu\nu}),
\end{equation}
where 
\begin{equation}
T^{(1)}_{\mu\nu}=\nb_{\mu}\vp\nb_{\nu}\vp-\frac{1}{2}g_{\mu\nu}\nb^{\lambda}\vp\nb_{\lambda}\vp,
\end{equation}
\begin{eqnarray}
\nonumber T^{(2)}_{\mu\nu}=\frac{1}{2}\nb_{\mu}\vp\nb_{\nu}\vp R-2\nb^{\lambda}\vp\nb_{\nu}\vp R_{\lambda\mu}+\frac{1}{2}\nb^{\lambda}\vp\nb_{\lambda}\vp G_{\mu\nu}-g_{\mu\nu}\left(-\frac{1}{2}\nb_{\lambda}\nb_{\kappa}\vp\nb^{\lambda}\nb^{\kappa}\vp\right.\\\left.+\frac{1}{2}(\nb^2\vp)^2-R_{\lambda\kappa}\nb^{\lambda}\vp\nb^{\kappa}\vp\right)
-\nb_{\mu}\nb^{\lambda}\vp\nb_{\nu}\nb_{\lambda}\vp+
\nb_{\mu}\nb_{\nu}\vp\nb^2\vp-R_{\lambda\mu\kappa\nu}\nb^{\lambda}\vp\nb^{\kappa}\vp
\end{eqnarray}
As it is easy to see the term $T^{(1)}_{\mu\nu}$ is the ordinary energy momentum tensor for the scalar field $\vp$ and the second term $T^{(2)}_{\mu\nu}$ appears due to  nonminimal derivative coupling of the scalar field with gravity. The equations for the scalar field takes the following form:
\begin{equation}\label{scal_f_eq}
(\alpha g_{\mu\nu}-\eta G_{\mu\nu})\nb^{\mu}\nb^{\nu}\vp=0,
\end{equation}

We are to obtain a $(n+1)$-dimensional slowly rotating black hole solution. It is known that when $n>3$ a black hole might have several angular momenta which represent rotations in different nonintersecting planes. In this paper we consider the simplest case of rotation in one plane, so we would have the only parameter $a$ characterizing the rotation. The metric is supposed to take the form:
\begin{equation}\label{metric}
ds^2=-U(r)dt^2+W(r)dr^2-2af(r)\sin^2{\th}dtd\chi+r^2(d\th^2
+\sin^2{\th}d\chi^2+\cos^2{\th}d\Omega^2_{n-3}),
\end{equation}  
where $a$ is the parameter related to the angular momentum of the black hole and $d\Omega^2_{n-3}$ is the standard metric of a unit $(n-3)$-dimensional sphere. It should be noted that all the unknown functions $U(r)$, $W(r)$ and $f(r)$ depend on the radial coordinate $r$ only.

Having used the evident from of the metric (\ref{metric}) one might write the Einstein equations  (\ref{eins_eq}) in the form:
\begin{equation}
\label{eq_1}\frac{r}{2}\left(2W+\frac{3}{2}\e(\vp\rq{})^2\right)\left(\frac{U\rq{}}{U}-\frac{W\rq{}}{W}\right)=-(n-2)\e(\vp\rq{})^2+2(n-2)W(W-1)-\e r\vp\rq{}\rq{}\vp\rq{}-\frac{4\L}{n-1}r^2W^2,
\end{equation}
\begin{equation}\label{eq_2}
r\left(2W+\frac{3}{2}\e(\vp\rq{})^2\right)\frac{U\rq{}}{U}=\frac{\al}{n-1}r^2W(\vp\rq{})^2+2(n-2)W(W-1)+\frac{(n-2)}{2}\e(\vp\rq{})^2(W-3)-\frac{4\L}{n-1}r^2W^2,\end{equation}
\begin{eqnarray}\label{eq_3} \nonumber\left(1-\e\frac{(\vp\rq{})^2}{4 W}\right)\left[\frac{1}{2UW}\left(U\rq{}\rq{}-\frac{(U\rq{})^2}{2U}-\frac{U\rq{}W\rq{}}{2W}\right)+\frac{n-2}{2rW}\left(\frac{U\rq{}}{U}-\frac{W\rq{}}{W}\right)-\right.\\\nonumber\left.\frac{(n-2)(n-3)}{2r^2W}(W-1)\right]=-\frac{\al}{4W}(\vp\rq{})^2+\frac{\e}{2W^2}\left(-\vp\rq{}\rq{}\vp\rq{}\left(\frac{U\rq{}}{2U}+\frac{n-2}{r}\right)-\right.\\\left.\frac{(n-2)}{2r}(\vp\rq{})^2\left(\frac{U\rq{}}{U}-2\frac{W\rq{}}{W}+\frac{n-3}{r}\right)-\frac{(\vp\rq{})^2}{2U}\left(U\rq{}\rq{}-\frac{(U\rq{})^2}{2U}-\frac{U\rq{}W\rq{}}{W}\right)\right),
\end{eqnarray}
\begin{eqnarray}\label{eq_4}
\left(1+\e\frac{(\vp\rq{})^2}{4W}\right)X\rq{}=\left(1+\e\frac{(\vp\rq{})^2}{4W}\right)\left(\frac{U\rq{}}{2U}+\frac{W\rq{}}{2W}-\frac{n-1}{r}\right)X-\frac{\e}{2W}\left(\vp\rq{}\rq{}-\frac{(\vp\rq{})^2W\rq{}}{2W^2}\right)X,
\end{eqnarray}
and here $X=f\rq{}-2{f}/{r}$.

The equation (\ref{scal_f_eq}) can be represented in the form:
\begin{equation}\label{current_diff}
\frac{d}{dr}\left(\sqrt{\frac{U}{W}}r^{n-1}\left[\al-\e\frac{(n-1)}{2rW}\left(\frac{U\rq{}}{U}-\frac{(n-2)}{r}(W-1)\right)\right]\vp\rq{}\right)=0
\end{equation}
From the latter equation it follows immediately that:
\begin{equation}\label{current_int}
\sqrt{\frac{U}{W}}r^{n-1}\left[\al-\e\frac{(n-1)}{2rW}\left(\frac{U\rq{}}{U}-\frac{(n-2)}{r}(W-1)\right)\right]\vp\rq{}=C
\end{equation}
To simplify integration of the equations (\ref{eq_1})-(\ref{eq_4}) we can choose the constant $C$ to be equal to zero ($C=0$). This choice allows one to  decouple the system of equations (\ref{eq_1})-(\ref{eq_2}) easily. It is worth noting that the latter choice is equivalent to the  condition:
\begin{equation}\label{metr_cond}
\al g^{rr}-\e G^{rr}=0.
\end{equation}
So when one imposes the condition (\ref{metr_cond}) the field equation (\ref{scal_f_eq}) is satisfied immediately. We also remark that the same requirement was imposed on the component of metric tensor and corresponding component of the Einstein tensor in the works where black holes with nonminimally coupled scalar field were considered \cite{Minamitsuji_PRD2014,Kobayashi_PTEP14}.

It should be noted that the metric (\ref{metric}) contains three unknown functions, whereas Einstein equations (\ref{eins_eq}) give rise to the four written above equations (\ref{eq_1})-(\ref{eq_4}) so one of them appears to be a ``redundant'' equation. In our case we can take the system of the equations (\ref{eq_1})-(\ref{eq_2}) which together with the condition (\ref{metr_cond}) allow us to find the unknown functions $U(r)$, $W(r)$ and $\vp\rq{}$.  Usually the ``redundant'' equations are used for obtaining of some constraints on integration constants that appears due to integration of the independent system of equations. One can verify that in our case the ``redundant'' equation (\ref{eq_3}) will be just an identity for the solutions that we will obtain. Finally, the equation (\ref{eq_4}) is used for the purpose of finding the function $f(r)$.

\section{Solutions of the field equations} 
The solution we are going to obtain will depend on the sign of parameters $\al$ and $\e$. Firstly, we investigate the solution for the case $\al>0$ and $\e>0$. It should be remarked that the solution depends on the parity of dimension of space $n$. In case of the odd $n$ we have:
\begin{equation}\label{funct_u_odd}
U(r)=1-\frac{\mu}{r^{n-2}}-\frac{2\L}{n(n-1)}r^2+\frac{(\al+\L\e)^2}{2\al\e(n-1)}\left[(-1)^{\frac{n+1}{2}}\frac{d^n}{r^{n-2}}\arctan\left({\frac{r}{d}}\right)+\sum^{\frac{n-1}{2}}_{j=0}(-1)^jd^{2j}\frac{r^{2(1-j)}}{n-2j}\right],
\end{equation} 
and here $d^2=\e(n-1)(n-2)/2\al$ and $\mu$ is a constant of integration which is related to mass of the black hole (so called mass parameter).

For the case of even $n$ we obtain:
\begin{equation}\label{funct_u_even}
U(r)=1-\frac{\mu}{r^{n-2}}-\frac{2\L}{n(n-1)}r^2+\frac{(\al+\L\e)^2}{2\al\e(n-1)}\left[(-1)^{\frac{n}{2}}\frac{d^n}{2r^{n-2}}\ln\left(\frac{r^2}{d^2}+1\right)+\sum^{\frac{n}{2}-1}_{j=0}(-1)^jd^{2j}\frac{r^{2(1-j)}}{n-2j}\right],
\end{equation}
It is worth emphasizing that when $\mu=0$ nontrivial solution of Einstein equations still exists and it represents so called gravitational solitons. 

Two other functions, namely $W(r)$ and $\vp\rq{}(r)$, for the even as well as for the odd dimensions take the form as follows:
 \begin{eqnarray}
\label{funct_W} W(r)=\frac{((\al-\L\e)r^2+\e(n-1)(n-2))^2}{(2\al r^2+\e(n-1)(n-2))^2U(r)},\\\label{pot_phi} (\vp\rq{})^2=-\frac{4(\al+\L\e)r^2W(r)}{\e(2\al r^2+\e(n-1)(n-2))}
 \end{eqnarray} 
 It is worth being noted that the square of derivative of the scalar potential (\ref{pot_phi}) has to be nonnegative outside the black hole\rq{}s horizon to provide nonnegativity of the kinetic energy of the scalar field $\nabla_{\mu}\vp\nabla^{\mu}\vp$ in the outer domain and due to the change of sign of the metric function $W(r)$ when one crosses the horizon the kinetic energy of the scalar field remains positive definite in the inner domain. The positivity can be provided when the cosmological constant is negative and satisfies the condition $\L<-\e/\al$.
  
 Finally, the function $f(r)$  which allows to take into account slow rotations takes the form:
 \begin{equation}
 f(r)=C_2r^{2-n}+C_3r^2,
 \end{equation}
 where $C_2$ and $C_3$ are integration constants. It should be stressed that the function $f(r)$ has completely the same dependence of the radial coordinate $r$ as it is in the case of a slowly rotating Kerr-AdS(-dS) solution in standard General Relativity.
 
 Now we consider asymptotic behaviour of our metric functions for different regimes. Firstly we examine the odd dimensional case and then proceed to the even one. 
 
Firstly we investigate the behaviour of the metric functions $U(r)$ and $W(r)$ for small $r$. To perform that task we decompose the function $\arctan(r/d)$ into a series for small $r$. It is easy to verify that after the decomposition we obtain:
\begin{equation}\label{u_small_r}
U(r)=1-\frac{2\L}{n(n-1)}r^2-\frac{\mu}{r^{n-2}}+\frac{(\al+\L\e)^2}{\e^2(n-1)^2(n^2-4)}r^4+{\cal O}(r^6)
\end{equation} 
Here the first three terms completely recover the Schwarzschild-AdS solution and the last two terms represent the correction due to nonminimal derivative coupling. For the other metric function $W(r)$ we reveal that $W(r)\simeq 1/U(r)$, so here we go back again to the relation between the metric functions of the Schwarzschild-AdS black hole\rq{}s solution. The result given by the relation (\ref{u_small_r}) has clear explanation, namely the coupling with the scalar field and in particular nonminimal coupling has to some extent cosmological origin, so it affects on the behaviour of the solution on larger, cosmological scales. Since we consider a black hole which is supposed to be a relatively compact object, the modification of solutions obtained in the framework of standard General Relativity is rather moderate and this fact is reflected by the relation (\ref{u_small_r}).

Now we consider the regime of large nonminimal coupling (large $\e$), so we suppose that the term which represents the minimal coupling, namely the term which contains the factor $\al$  is substantially smaller than the term which represents the nonminimal coupling (the trem with the factor $\e$). In this case we have:
\begin{equation}\label{u_large_e}
U(r)=1-\frac{2\L}{n(n-1)}r^2-\frac{\mu}{r^{n-2}}+\frac{\L^2}{(n-1)^2(n^2-4)}r^4+{\cal O}\left(\frac{1}{\e}\right).
\end{equation}
It is worth remarking that:
\begin{equation}\label{rel_uw}
U(r)W(r)\simeq \left(1-\frac{\L}{(n-1)(n-2)}r^2\right)^2+{\cal O}\left(\frac{1}{\e}\right).
\end{equation}
For large distances ($r$ is large) an asymptotic relation for the $\arctan(r/d)$  can be utilized and as a consequence we arrive at the expression:
\begin{eqnarray}\label{as_larg_r}
\nonumber U(r)=\frac{(2\al+\L\e)(\al-\L\e)}{4\al^2}+\frac{(\al-\L\e)^2}{2n(n-1)\al\e}r^2+\frac{(\al+\L\e)^2}{2(n-1)\al\e}\sum^{(n-1)/2}_{j=2}(-1)^jd^{2j}\times\\\frac{r^{2(1-j)}}{n-2j}+\left(-\mu+(-1)^{(n+1)/2}\frac{\pi(\al+\L\e)^2}{4(n-1)\al\e}d^n\right)\frac{1}{r^{n-2}}+{\cal O}\left(\frac{1}{r^{n-1}}\right).
\end{eqnarray}
We point out here that the dominating term at the infinity is of the order $\sim r^2$ so the behaviour of the metric at the infinity is the same as for AdS-type black holes, but in contrast to standard General Relativity here we have ``effective'' cosmological constant $(\al-\L\e)^2/4\al\e$ instead of its ``pure'' value $\L$ in the ordinary case. We also note, that in the case of large $r$ for the product of the metric functions we have: $U(r)W(r)\simeq (\al-\L\e)^2/4\al^2$. The second peculiarity of the written asymptotic relation for the metric function (\ref{as_larg_r}) is the presence of the terms combined in the sum which decrease at the infinity slowly than the term $\sim 1/r^{n-2}$, there are no corresponding terms in the ordinary case. The term $1/r^{n-2}$ represents Schwarzschild-like behaviour  but instead of the constant $\mu$ a new factor appears which depends on the parameters $\al$ and $\e$  as well as on the cosmological constant $\L$. It is also worth being remarked that when $n=3$ we do not have the terms that decrease slowly than $1/r$, so Schwarzschild-like behaviour of decaying terms can be recovered exactly only in this case. 

We remark that when $\L=-\al/\e$ we obtain exact Schwarzschild-AdS(-dS) solution. So, one can write:
\begin{equation}\label{Sch_AdS_exact}
U(r)=1-\frac{2\L}{n(n-1)}r^2-\frac{\mu}{r^{n-2}}.
\end{equation} 
Here we also have that $W(r)U(r)=1$ and $\vp\rq{}=0$ so the scalar potential might be only a constant and this fact explains why we arrive at the exact solution (\ref{Sch_AdS_exact}). We also remark that the identical situation takes place for the case of even dimensions when $\L=-\al/\e$. 

Now we consider the case of even space dimensions, so we analyze the metric function (\ref{funct_u_even}). To study the behaviour of the metric function $U(r)$ (\ref{funct_u_even}) for small $r$ we decompose the logarithmic term in it into a series. Fortunately enough the first several terms in this decomposition allow to eliminate the sum in the metric function (\ref{funct_u_even}) completely, so the situation is identical to the case of the odd $n$ that we had before, so for the small distances we arrive at the relation (\ref{u_small_r}) again. Thus, for small distances there is no difference between the cases of different parities of $n$ as it should be, because we recover the standard Schwarzschild-AdS solution which is identical for the both parities of $n$. Similarly we obtain the asymptotes for the case when nonminimal coupling is large ($\e$ is large). Finally, we investigate the metric function (\ref{funct_u_even}) when $r$ is considerably greater than the parameter $d$. We can use asymptotic formula for the logarithm that we have in the function $U(r)$:
\begin{equation}
\ln\left(\frac{r^2}{d^2}+1\right)=2\ln\left(\frac{r}{d}\right)+\sum^{+\infty}_{j=0}(-1)^j\frac{d^{2(j+1)}}{(j+1)r^{2(j+1)}}
\end{equation} 
Having used the latter relation we can write:
\begin{eqnarray}\label{as_larg_r_even}
\nonumber U(r)=\frac{(2\al+\L\e)(\al-\L\e)}{4\al^2}+\frac{(\al-\L\e)^2}{2n(n-1)\al\e}r^2+\frac{(\al+\L\e)^2}{2(n-1)\al\e}\sum^{n/2-1}_{j=2}(-1)^jd^{2j}\frac{r^{2(1-j)}}{n-2j}+\\
\left(-\mu+(-1)^{n/2}\frac{(\al+\L\e)^2}{2(n-1)\al\e}d^n\ln\left(\frac{r}{d}\right)\right)\frac{1}{r^{n-2}}+{\cal O}\left(\frac{1}{r^n}\right).
\end{eqnarray} 
The written above relation demonstrates some similarity with the given before relation for the asymptotic of the metric function $U(r)$ for odd $n$ (\ref{as_larg_r}). First of all, for large $r$ the metric (\ref{funct_u_even}) shows AdS-like behaviour similarly as it was for the metric (\ref{funct_u_odd}) and it supports the fact that the large $r$ asymptotic is supposed to be the same for both parities of $n$. The second similarity between the relations (\ref{as_larg_r}) and (\ref{as_larg_r_even}) is due to the fact that for large $n$ both of them contain the sum of the terms decaying slowly than $1/r^{n-2}$ at infinity. The main difference between the relations (\ref{as_larg_r}) and (\ref{as_larg_r_even}) is due to logarithmic factor $\ln(r/a)$ which makes the decay of the $\sim 1/r^{n-2}$ term slower at infinity for even $n$ than it was for the odd $n$. The less notable difference  between the mentioned asymptotes is related to the terms that go down faster than $1/r^{n-2}$ at infinity. For the case of odd $n$  they have the leading term of the order $\sim 1/r^{n-1}$ while for the even $n$ the leading term is $\sim 1/r^n$.

\begin{figure}
\centerline{\includegraphics[scale=0.33,clip]{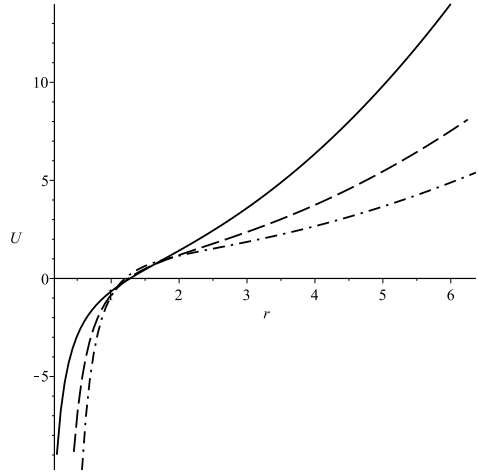}}
\caption{The dependence $U(r)$ for different values of dimensions. The solid, dashed and dash-dotted curves correspond to $n=3$, $n=4$ and $n=5$ respectively. The other parameters are held fixed, namely $\al=0.2$, $\e=0.4$ and $\L=-1$.}\label{metr_f_1}
\end{figure}

The behaviour of the metric function $U(r)$ is demonstrated on the Fig. (\ref{metr_f_1}). This figure shows that the metric function $U(r)$ is monotonically increasing with the only root $r_+$  corresponding to the horizon point. The increase of this function for large $r$ becomes slower with increasing of the dimension $n$ what is seen easily from the relations (\ref{funct_u_odd}) and (\ref{funct_u_even}).  

The important characteristics which can explain whether the obtained solution represents the black hole is the behaviour of Kretschmann scalar. It can be written in the form:
\begin{equation}\label{Kr_scalar}
R_{\mu\nu\k\l}R^{\mu\nu\k\l}=\frac{1}{U^2W^2}\left(U\rq{}\rq{}-\frac{(U\rq{})^2}{2U}-\frac{U\rq{}W\rq{}}{2W}\right)^2+\frac{(n-1)}{r^2W^2}\left(\frac{(U\rq{})^2}{U^2}+\frac{(W\rq{})^2}{W^2}\right)+2(n-1)(n-2)\frac{(W-1)^2}{r^4W^2}
\end{equation}
It should be noted that in the latter relation we have neglected the terms proportional to $a^2$. For simplicity we have not substituted the evident form of the functions $U(r)$ and $W(r)$. Special attention should be paid to the behaviour of Kretschmann scalar $R_{\mu\nu\k\l}R^{\mu\nu\k\l}$ at the points where the metric shows singular behaviour. Firstly, it can be shown that at the horizon the Kretschmann scalar is nonsingular which tells us that at this point we have ordinary coordinate singularity that is typical for any other black hole. When the $r\rightarrow 0$ for the Kretschmann scalar we have:
\begin{equation}\label{Kr_sc_zero}
R_{\mu\nu\k\l}R^{\mu\nu\k\l}\sim \frac{n(n-1)^2(n-2)\mu^2}{r^{2n}}
\end{equation} 
and one can conclude that in this case we have physical singularity. We remark that due to the fact that when $r\rightarrow 0$ the leading terms of the metric functions (\ref{funct_u_odd}) and (\ref{funct_u_even}) are those that give Schwarzschild-like behaviour and the Kretschmann scalar (\ref{Kr_sc_zero}) at small distances close the singularity point ($r=0$)  shows the same dependence of $r$ as it is for ordinary Schwarzschild solution.
In the case when $r\rightarrow\infty$ one obtains:
 \begin{equation}\label{KR_sc_inf}
R_{\mu\nu\k\l}R^{\mu\nu\k\l}\sim \frac{8(n+1)\al^2}{n(n-1)^2\e^2}
\end{equation} 
So at the infinity the Kretschmann scalar gets finite value which depends on the dimension of space as well as on the parameters $\al$ and $\e$ which are given in the action (\ref{action}) and does not depend on the constant of integration that characterizes a particular solution of equations of motion. We also note that for large distances ($r\rightarrow\infty$) the Kretschamnn scalar takes the same dependence of the dimension of space $n$ as it takes for AdS-solution and it can be treated as an additional fact confirming the AdS-behaviour of the obtained solution for large  distances that has been mentioned earlier.  The given analysis shows that the only physical singularity appears when $r=0$ and it corroborates the fact that we have black hole solution.   
  
\begin{figure}
\centerline{\includegraphics[scale=0.33,clip]{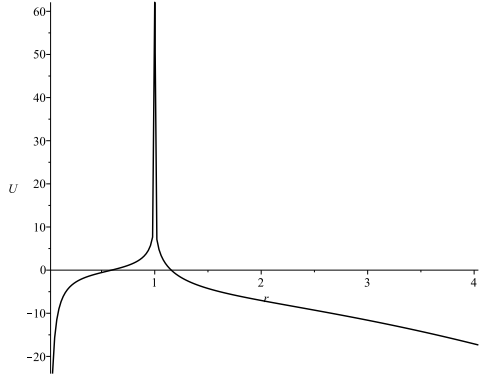}}
\caption{Function $U(r)$ given by the relation (\ref{funct_u_odd_neg}) for negative value of parameter $\e$. The fixed parameters are taken as follows: $n=3$, $\al=0.2$, $\e=0.2$ and $\L=-1$.}\label{f_u_neg}
\end{figure}

 For the case of negative relation $\e/\al<0$ (in the following it is supposed that $\al>0$ and $\e<0$) we have other expressions for the metric function $U(r)$. Again we treat separately the cases of odd and even dimensions of the space $n$. In the case of negative $\e$ the   expressions (\ref{funct_W}) and (\ref{pot_phi}) are valid and it means that to provide the existence of stable solutions we have to impose that $\al/|\e|>\L$, so in this case the cosmological constant might be positive as well as negative in contrast to the situation that we had before where the only negative values for the cosmological constant were allowed. It should also be stressed that the point $r=|\e|(n-1)(n-2)/2\al$ is the point where the function $(\vp\rq{})^2$  is divergent and crossing of this point changes its sign, so it separates the stable and unstable domains. Now we can write a solution for the odd $n$:
\begin{equation}\label{funct_u_odd_neg}
U(r)=1-\frac{\mu}{r^{n-2}}-\frac{2\L}{n(n-1)}r^2+\frac{(\al+\L\e)^2}{2\al\e(n-1)}\left[\frac{d^n}{2r^{n-2}}\ln\left|{\frac{r-d}{r+d}}\right|+\sum^{(n-1)/2}_{j=0}d^{2j}\frac{r^{2(1-j)}}{n-2j}\right],
\end{equation}  
It should be noted that here $d^2=|\e|(n-1)(n-2)/2\al$. One can see that the logarithmic term of the expression (\ref{funct_u_odd_neg}) is divergent when $r=d$, so we have some kind of coordinate singularity at this point. This singularity can be explained by the mentioned above fact about the divergence of the derivative of the scalar potential at this point, but for the function $U(r)$ we do not have the change of sign in the vicinity of the point. We also remark  that the Kretschmann scalar  $R_{\kappa\lambda\mu\nu}R^{\kappa\lambda\mu\nu}$ does not diverge at the point $r=d$. It can be shown that  the function $U(r)$ might have two roots ($U(r)=0$), namely the first of them lies below the point $r_d=d$ ($r_i<r_d$), whereas the other one $r_{c}$ is greater then $r_d$ ($U(r_c)=0$) and the function (\ref{funct_u_odd_neg}) becomes negative for larger values of $r$ ($r>r_c$) (see the Fig.(\ref{f_u_neg})). It leads to the conclusion that the point $r=r_c$ can be treated as a cosmological horizon. This analysis bring us to the conclusion that the smaller root $r_i$ should be identified with the event horizon of the black hole, but as it has already been mentioned the point of instability $r_d$ where the kinetic energy of the scalar field $\vp$ changes sign appears to be in the outside domain of a black hole. So, we can conclude that the obtained solution (\ref{funct_u_odd_neg}) cannot be treated as a black hole. We also point out that with increasing of the cosmological constant $\L$, but when still $\L<\al/|\e|$ the metric function (\ref{funct_u_odd_neg}) would be negative for almost all values of $r$ excluding the point $r_d$. This fact additionally corroborate the conclusion that we cannot consider the obtained solution for metric function $U(r)$ (\ref{funct_u_odd_neg}) as metric function for a black hole.

Finally, we obtain the solution for even $n$:
\begin{equation}\label{f_u_neg_even}
U(r)=1-\frac{\mu}{r^{n-2}}-\frac{2\L}{n(n-1)}r^2+\frac{(\al+\L\e)^2}{2\al\e(n-1)}\left[\frac{d^n}{2r^{n-2}}\ln\left|\frac{r^2}{d^2}-1\right|+\sum^{n/2-1}_{j=0}d^{2j}\frac{r^{2(1-j)}}{n-2j}\right].
\end{equation}
The careful analysis of the obtained metric function (\ref{f_u_neg_even}) shows that its behaviour is completely identical to the behaviour of previously examined function (\ref{funct_u_odd_neg}) so it immediately gives rise to the conclusion that the obtained solution (\ref{f_u_neg_even}) does not represent a black hole.  To sum it up one can conclude that the black hole solutions appear only for positive parameter $\e$.   

\begin{figure}
\centerline{\includegraphics[scale=0.33,clip]{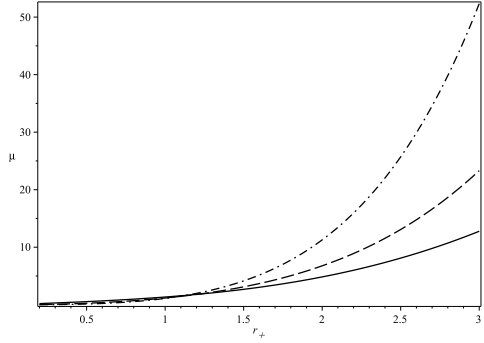}}
\caption{Mass parameter $\mu$ as a function of horizon radius $r_+$ for several values of dimension of space. The solid, dashed  and dash-dotted curves correspond to $n=3$, $n=4$ and $n=5$ respectively. The other parameters are equal to: $\al=0.2$, $\e=0.4$ and $\L=-1$.}\label{mass_param}
\end{figure}

Having used the relations (\ref{funct_u_odd}) and (\ref{funct_u_even}) one can obtain relations for the mass parameter $\mu$ as functions of horizon radius $r_+$. It can be written as follows:
\begin{equation}
\mu=\left(1-\frac{(\al+\L\e)^2}{4\al^2}\right)r^{n-2}_{+}+\frac{(\al-\L\e)^2}{2n(n-1)\al\e}r^n_{+}+\frac{(\al+\L\e)^2}{2(n-1)\al\e}\left[(-1)^{\frac{n+1}{2}}d^n\arctan\left(\frac{r_+}{d}\right)+\sum^{\frac{n-1}{2}}_{j=2}(-1)^jd^{2j}\frac{r^{n-2j}_{+}}{n-2j}\right]
\end{equation}
for odd $n$ and
\begin{equation}
\mu=\left(1-\frac{(\al+\L\e)^2}{4\al^2}\right)r^{n-2}_{+}+\frac{(\al-\L\e)^2}{2n(n-1)\al\e}r^n_{+}+\frac{(\al+\L\e)^2}{2(n-1)\al\e}\left[(-1)^{\frac{n}{2}}\frac{d^n}{2}\ln\left(\frac{r^2_+}{d^2}+1\right)+\sum^{\frac{n}{2}-1}_{j=2}(-1)^jd^{2j}\frac{r^{n-2j}_{+}}{n-2j}\right]
\end{equation}
for the case of even $n$. The dependence $\mu=\mu(r_+)$ for different dimensions $n$ is depicted on the Fig.(\ref{mass_param}). The Figure shows that the mass parameter is monotonically increasing function of the horizon radius $r_+$ and for larger values of $r_+$ the rise of this function is faster for higher values of $n$. In the following we will show that the mass of the black hole $M$ is directly proportional to the mass parameter $\mu$ so described above the behaviour of the mass parameter can be completely applied to the black hole's mass. It should also be noted that the monotonous behaviour of  the mass parameter that has just been mentioned makes the thermodynamical functions such as internal energy or enthalpy to be well defined which is extremely important for thermodynamics of the black hole.

\section{Thermodynamics of the black hole}
\subsection{Temperature of the black hole}
\begin{figure}
\centerline{\includegraphics[scale=0.33,clip]{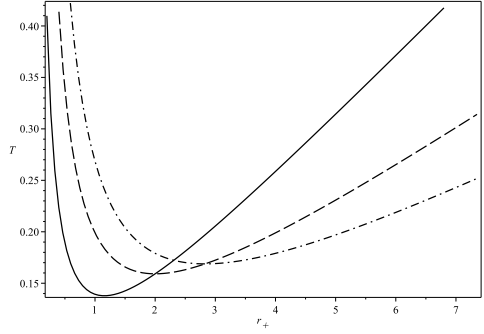}}
\caption{Temperature of the black hole as a function of horizon radius $r_+$ for several values of dimension of space. The solid, dashed  and dash-dotted curves correspond to $n=3$, $n=4$ and $n=5$ respectively. The other parameters are equal to: $\al=0.2$, $\e=0.4$ and $\L=-1$.}\label{Temp_BH}
\end{figure}

In this section we will obtain thermodynamic functions for the given  above black hole solution and investigate their behaviour. It is known that all black holes possess such characteristic as a temperature. To obtain the black hole\rq{}s temperature we rely on a well known technique which uses the notion of surface gravity $\kappa$:
\begin{equation}\label{surf_grav}
\kappa^2=-\frac{1}{2}\nabla_{a}\chi_b\nabla^{a}\chi^{b},
\end{equation}
where $\chi^a$ are components of a Killing vector which should be null on the event horizon. It is easy to verify that the time translation vector $\chi^{\mu}=\partial/\partial t$ would be the very same Killing vector which satisfies mentioned above conditions in the first order approximation over the angular momentum parameter $a$. Having calculated the surface gravity by using the formula (\ref{surf_grav}) and taking into account the definition of black hole\rq{}s temperature we obtain:
\begin{equation}\label{temp_gen}
T=\frac{\kappa}{2\pi}=\frac{1}{4\pi}\frac{U\rq{}(r_+)}{\sqrt{U(r_+)W(r_+)}}
\end{equation}  
where $r_+$ denotes the event horizon radius of the black hole. It is clear that  because we have different evident forms for the metric functions $U(r)$ corresponding to some particular choice of the space dimension (even or odd $n$) we arrive at different expressions for the temperature which will be treated separately in the following.  
As it was noted above the black hole solution takes place when both parameters $\al$ and $\e$ are positive and as a result the temperature (\ref{temp_gen}) takes the following form:
\begin{eqnarray}\label{T_odd}
\nonumber T=\frac{1}{4\pi}\frac{2\al r^2_{+}+\e(n-1)(n-2)}{(\al-\L\e)r^2_{+}+\e(n-1)(n-2)}\left(\frac{(\al-\L\e)^2}{2(n-1)\al\e}r_{+}+\left(1-\frac{(\al+\L\e)^2}{4\al^2}\right)\times\right.\\\left.\frac{(n-2)}{r_+}+\frac{(\al+\L\e)^2}{2(n-1)\al\e}\left[(-1)^{(n+1)/2}\frac{d^{n+1}}{r^{n-2}_{+}(r^2_{+}+d^2)}+\sum^{(n-1)/2}_{j=2}(-1)^jd^{2j}r^{1-2j}_{+}\right]\right)
\end{eqnarray}
for odd $n$ and
\begin{eqnarray}\label{T_even}
\nonumber T=\frac{1}{4\pi}\frac{2\al r^2_{+}+\e(n-1)(n-2)}{(\al-\L\e)r^2_{+}+\e(n-1)(n-2)}\left(\frac{(\al-\L\e)^2}{2(n-1)\al\e}r_{+}+\left(1-\frac{(\al+\L\e)^2}{4\al^2}\right)\times\right.\\\left.\frac{(n-2)}{r_+}+\frac{(\al+\L\e)^2}{2(n-1)\al\e}\left[(-1)^{n/2}\frac{d^{n}}{r^{n-3}_{+}(r^2_{+}+d^2)}+\sum^{n/2-1}_{j=2}(-1)^jd^{2j}r^{1-2j}_{+}\right]\right)
\end{eqnarray}
for even $n$. It is worth mentioning that the temperature in both written above relations is represented as a function of the horizon radius. These dependences for different dimensions $n$ are shown on the Fig. (\ref{Temp_BH}). As it is easy to see that qualitatively the curves are similar with some specific minimum point which reflects the existence of the Hawking-Page phase transition \cite{Hawking_CMP83}. One can also conclude that with the increasing of $n$, the point of minimum goes to larger radii $r_+$ and the increase of the temperature for small $r_+$ becomes faster whereas for large $r_+$ it goes up slowly than for smaller $n$.
\subsection{Wald formalism and black hole\rq{}s entropy}
In order to derive the generalized first law of black hole thermodynamics Wald proposed an elegant formalism which allows to obtain variation of a Hamiltonian with help of a Noether conserved current \cite{Wald_PRD93,Iyer_PRD94}. It should be noted that Wald formalism is applicable for quite  general diffeomorphism-invariant theories with higher derivatives. We also note that this formalism was used to derive the first law of thermodynamics for numerous black holes with AdS-asymptotic behaviour, namely for Einstein-scalar \cite{Liu_PLB14, Lu_JHEP15}, Einstein-Proca \cite{Liu_JHEP14}, Einstein-Yang-Mills theories \cite{Fan_JHEP15} and some other variants \cite{Fan_PRD15,Liu_JHEP14_2}. Since we depart from the same action that was investigated in the paper \cite{Feng_JHEP15} so our following derivations are analogous to those ones in the mentioned work. 
Variating the action (\ref{action}) we obtain the surface terms of the form:
\begin{eqnarray}\label{surf_term}
\nonumber J^{\mu}=2\frac{\partial L}{\partial R_{\kappa\lambda\mu\nu}}\nabla_{\lambda}\delta g_{\kappa\nu}-2\nabla_{\nu}\frac{\partial L}{\partial R_{\kappa\mu\nu\lambda}}\delta g_{\kappa\lambda}+\frac{\partial L}{\partial(\nabla_{\mu}\vp)}\delta\vp\\=J^{\mu}_g+\al J^{\mu}_{\vp}+\eta\left(J^{\mu}_{g1}+J^{\mu}_{\vp 1}\right),
\end{eqnarray}
and here we use the following notations:
\begin{equation}
J^{\mu}_g=g^{\mu\nu}g^{\kappa\lambda}\nabla_{\lambda}
\left(\delta g_{\nu\kappa}\right)-g^{\mu\kappa}g^{\lambda\nu}\nabla_{\kappa}\left(\delta g_{\lambda\nu}\right), \quad J^{\mu}_{\vp}=-g^{\mu\lambda}\nabla_{\lambda}\vp\delta\vp, \quad J^{\mu}_{\vp 1}=G^{\mu\lambda}\nabla_{\lambda}\vp\delta\vp
\end{equation}
\begin{eqnarray}
\nonumber J^{\mu}_{g1}=-\frac{1}{4}(\nabla\vp)^2J^{\mu}_g+\frac{1}{2}g^{\mu\nu}\nabla^{\kappa}\vp\nabla^{\lambda}\vp
\nabla_{\lambda}\left(\delta g_{\nu\kappa}\right)-\frac{1}{4}g^{\mu\nu}\nabla^{\kappa}\vp\nabla^{\lambda}\vp
\nabla_{\nu}\left(\delta g_{\lambda\kappa}\right)-\frac{1}{4}g^{\lambda\nu}\nabla^{\kappa}\vp\nabla^{\mu}\vp
\nabla_{\kappa}\left(\delta g_{\lambda\nu}\right)\\+\frac{1}{2}\left(g^{\mu\lambda}\nabla^{\nu}\nabla^{\kappa}\vp
\nabla_{\kappa}\vp-\nabla^{\mu}\vp\nabla^{\lambda}
\nabla^{\nu}\vp\right)\delta g_{\lambda\nu}+\frac{1}{4}\left(\nabla^{\mu}\vp\nabla^{\kappa}\nabla_{\kappa}\vp-
\nabla^{\mu}\nabla^{\kappa}\vp\nabla_{\kappa}\vp\right)
g^{\lambda\nu}\delta g_{\lambda\nu}
\end{eqnarray}

According to Wald procedure a 1-form can be defined $J_{(1)}=J_{\mu}dx^{\mu}$ and as a result its Hodge dual can be written:
\begin{equation}
\Theta_{(n)}=* J_{(1)}
\end{equation} 
Then it is supposed that an infinitesimal diffeomorphism is performed  $(\delta x^{\mu}=\xi^{\mu})$. As a result one can write:
\begin{equation}
J_{(n)}=\Theta_{(n)}-i_{\xi}*L_{(0)}=-d*J_{(2)},
\end{equation}
where it is assumed that the equations of motion are satisfied.   Here it is also supposed that $i_{\xi}$ denotes contraction of the infinitesimal field $\xi^{\mu}$ with the  first index of $n$-form. Now the form $*J_{(2)}$ can be identified with a $(n-1)$-form, namely $Q_{(n-1)}\equiv*J_{(2)}$. To install  the relation with the first law of black hole thermodynamics the infinitesimal vector $\xi^{\mu}$ should be chosen to  be time-like Killing vector which is null on the horizon. It was shown  that the variation of the Hamiltonian with respect to the integration constant of a specific solution can be represented by the following expression \cite{Wald_PRD93,Iyer_PRD94}:
\begin{equation}
\delta {\cal H}=\delta\int_{c}J_{(n)}-\int_{c}d\left(i_{\xi}\Theta_{(n)}\right)
=\int_{\Sigma^{n-1}}\delta Q_{(n-1)}-i_{\xi}\Theta_{(n)}
\end{equation}  
and here $c$ denotes a Cauchy surface and $\Sigma^{n-1}$  is its boundary with two components, namely at the horizon and at the infinity. Wald formalism demonstrates that the first law of thermodynamics can be obtained from the relation:
\begin{equation}
\delta {\cal H}_{\infty}=\delta {\cal H}_{+}
\end{equation} 
It was shown that for the theories with nonminimal derivative coupling \cite{Feng_JHEP15} one arrives at:
\begin{eqnarray}
\nonumber J_{\mu_1...\mu_n}=e.o.m.+2\varepsilon_{\mu_1...\mu_n\nu}\nabla_{\lambda}\left(\nabla^{[\lambda}\xi^{\nu]}
-\frac{\eta}{4}(\nabla\vp)^2\nabla^{[\lambda}\xi^{\nu]}+\frac{\eta}{2}\nabla^{\kappa}\vp\nabla^{[\lambda}\vp\nabla_{\kappa}\xi^{\nu]}\right.\\+\frac{\eta}{2}\nabla^{[\lambda}(\nabla\vp)^2\xi^{\nu]}-\frac{\eta}{2}\nabla^{\kappa}(\nabla^{[\lambda}\vp\nabla_{\kappa}\vp)\xi^{\nu]}-\frac{\eta}{2}\nabla^{[\lambda}(\nabla ^{\nu]}\vp\nabla_{\kappa}\vp)\xi^{\kappa}\left.\right)
\end{eqnarray}
where $e.o.m.$ denotes the terms which give equations of motion.
As a consequence we arrive at:
\begin{eqnarray}
\nonumber Q_{\mu_1...\mu_{n-1}}=\varepsilon_{\mu_1...\mu_{n-1}\nu\lambda}\left(\frac{\partial L}{\partial R_{\nu\lambda\kappa\sigma}}\nabla_{\kappa}\xi_{\sigma}-2\xi_{[\sigma}
\nabla_{\kappa]}\left(\frac{\partial L}{\partial R_{\nu\lambda\kappa\sigma}}\right)\right)=\\
\nonumber \varepsilon_{\mu_1...\mu_n\nu\lambda}\left[\nabla^{\nu}\xi^{\lambda}-\frac{\eta}{4}(\nabla\vp)^{2}\nabla^{\nu}\xi^{\lambda}+\frac{\eta}{2}\nabla^{\kappa}\vp\nabla^{\nu}\vp\nabla_{\kappa}\xi^{\lambda}+
\right.\\\left.\frac{\eta}{2}\nabla^{\nu}(\nabla\vp)^2\xi^{\lambda}-\frac{\eta}{2}\nabla^{\kappa}(\nabla^{\nu}\vp\nabla_{\kappa}\vp)\xi^{\lambda}-\frac{\eta}{2}\nabla^{\nu}(\nabla^{\lambda}\vp\nabla_{\kappa}\vp)\xi^{\kappa}\right],
\end{eqnarray}
\begin{eqnarray}
(i_{\xi}\Theta)_{\mu_1...\mu_{n-1}}=\varepsilon_{\mu_1...\mu_{n-1}\nu\lambda}\left(2\frac{\partial L}{\partial R_{\kappa\rho\nu\sigma}}\nabla_{\rho}\delta g_{\kappa\sigma}-2\nabla_{\sigma}\frac{\partial L}{\partial R_{\kappa\nu\sigma\rho}}\delta g_{\kappa\rho}+\frac{\partial L}{\partial(\nabla_{\nu}\vp)}\delta\vp\right)\xi^{\lambda}
\end{eqnarray}
Having used the chosen representation of the metric (\ref{metric}) we might write the forms $Q$, the contracted form $i_{\xi}\Theta$ and their difference. Firstly, we represent the terms corresponding to pure Einsteinian gravity:
\begin{equation}\label{Q_E}
Q=\frac{r^{n-1}}{\sqrt{UW}}U\rq{}\Omega_{(n-1)};
\end{equation}
\begin{equation}\label{iT_E}
i_{\xi}\Theta=-r^{n-1}\sqrt{UW}\left(\frac{U\rq{}}{2UW}\left(\frac{\delta U}{U}+\frac{\delta W}{W}\right)-\frac{\delta U\rq{}}{UW}+\frac{(n-1)}{rW^2}\delta W-\frac{\alpha}{W}\vp\rq{}\delta\vp\right)\Omega_{(n-1)}
\end{equation}
\begin{equation}\label{diff_QiT_E}
\delta Q-i_{\xi}\Theta=r^{n-1}\sqrt{UW}\left(\frac{n-1}{rW^2}\delta W-\frac{\al}{W}\vp\rq{}\delta\vp\right)\Omega_{(n-1)},
\end{equation}
and here $\Omega_{(n-1)}$ denotes the $n-1$-form over angular variables. We note that given above  relations coincide with corresponding relations given for static black holes \cite{Liu_PLB14,Lu_JHEP15,Feng_JHEP15}. The offdiagonal  term $g_{t\vp}$ which is related to rotation would give the terms proportional to $a^2$ and  they are not taken into account in the relations (\ref{Q_E})-(\ref{diff_QiT_E}). Now we write the forms corresponding to the nonminimally coupled terms:
\begin{equation}\label{Q_N}
Q_{\eta}=-\frac{n-1}{2}\eta r^{n-2}\sqrt{UW}\frac{(\vp\rq{})^2}{W^2}\Omega_{(n-1)},
\end{equation} 
\begin{equation}\label{iT_N}
i_{\xi}\Theta_{\eta}=-\frac{n-1}{2}\eta r^{n-2}\sqrt{UW}\frac{1}{W^2}\left(\frac{(\vp\rq{})^2}{2U}\delta U+\left[\frac{U\rq{}}{U}+\frac{n-2}{r}(1-W)\right]\vp\rq{}\delta\vp\right)\Omega_{(n-1)}
\end{equation}
\begin{equation}\label{diff_QiT_N}
\delta Q_{\eta}-i_{\xi}\Theta_{\eta}=-\frac{n-1}{2}\eta r^{n-2}\sqrt{UW}\left(\frac{\delta(\vp\rq{})^2}{W^2}-\frac{3}{2}\frac{(\vp\rq{})^2}{W^3}\delta W-\frac{1}{W^2}\left[\frac{U\rq{}}{U}+\frac{n-2}{r}(1-W)\right]\vp\rq{}\delta\vp\right)\Omega_{(n-1)}
\end{equation}
Taking into consideration the relations (\ref{diff_QiT_E}) and (\ref{diff_QiT_N}) one can obtain total difference of variations:
\begin{eqnarray}\label{tot_diff}
\nonumber\delta Q-i_{\xi}\Theta=(n-1)r^{n-2}\sqrt{UW}\left[\frac{1}{W^2}\left(1+\frac{\eta(\vp\rq{})^2}{4W}\right)\delta W-\frac{\eta}{2}\delta\left(\frac{(\vp\rq{})^2}{W}\right)\right]\Omega_{(n-1)}\\=(n-1)r^{n-2}
\frac{\sqrt{UW}}{W^2}\left(1+\frac{\eta(\vp\rq{})^2}{4W}\right)\delta W\Omega_{(n-1)},
\end{eqnarray}
where we used the fact that $\delta\left((\vp\rq{})^2/W\right)=0$. For further calculations it is more convenient to redefine the function $W(r)$ in the following way: $W(r)=1/g(r)$. As a result the final form of the relation (\ref{tot_diff}) can be represented as follows:
\begin{equation}\label{var_fin}
\delta Q-i_{\xi}\Theta=-(n-1)r^{n-2}\sqrt{\frac{U}{g}}\left(1+\frac{\eta}{4}g(\vp\rq{})^2\right)\delta g\Omega_{(n-1)}
\end{equation}
Now it necessary to calculate the variation (\ref{var_fin}) at the horizon as well as at the infinity. At the infinity we obtain:
\begin{equation}
\delta {\cal H}_{\infty}=\delta M=\frac{1}{16\pi}\int \delta Q-i_{\xi}\Theta =\frac{(n-1)\omega_{n-1}}{16\pi}\delta\mu
\end{equation}
The latter relation shows that variation of the Hamiltonian at the infinity in case of the black hole with nonminimal derivative coupling is completely identical to the corresponding relation in Einstein\rq{}s theory. As a result, the mass of the black hole can be written as follows:
\begin{equation}\label{bh_mass}
M=\frac{(n-1)\omega_{n-1}}{16\pi}\mu,
\end{equation} 
which coincides with the mass of Schwarzschild-anti-de Sitter black hole in General Relativity. The variation of Hamiltonian at the horizon can be cast in the form:
\begin{equation}\label{TD_diff}
\delta {\cal H}_{+}=\frac{(n-1)\omega_{n-1}}{16\pi}U\rq{}(r_+)r^{n-2}_{+}\delta r_{+}=\sqrt{U(r_+)W(r_+)}T\delta\left(\frac{{\cal A}}{4}\right)=\left(1+\frac{\e}{4}\frac{(\vp\rq{})^2}{W}\Big|_{r_+}\right)T\delta\left(\frac{{\cal A}}{4}\right).
\end{equation}
where ${\cal A}=\omega_{n-1}r^{n-1}_+$ is the horizon area of the black hole. It is evident that in the latter equation the one quarter of horizon area cannot be identified with the  the entropy of the black hole because such identification gives rise to violation of one of the basic relation of thermodynamics. To cure this situation firstly it was supposed that in general the entropy is not equal to a quarter of area of horizon but the right hand side of the relation is equal to $T\delta S$ to keep the well established relation from black hole thermodynamics, but this way might lead to quite complicated relation between the area ${\cal A}$ and the entropy $S$  which is not universal and depends on the considered solution \cite{Feng_JHEP15}. The other difficulty related to that alternative expression for entropy is due to the fact that it might not allow to obtain Smarr relation for black hole\rq{}s mass, temperature and entropy in general case, whereas the naive assumption about the same expression for the entropy as in ordinary General Relativity might give a chance to derive this relation. To overcome those difficulties it was proposed that an additional \lq\lq{}charge\rq\rq{} related to the scalar field $\vp(r)$ should be introduced \cite{Feng_PRD15}. To obtain that \lq\lq{}charge\rq\rq{} we rewrite the relation (\ref{TD_diff}) in the form:
\begin{equation}\label{def_ent}
\left(1+\frac{\e}{4}\frac{(\vp\rq{})^2}{W}\Big|_{r_+}\right)T\delta\left(\frac{{\cal A}}{4}\right)=T\delta\left(\left[1+\frac{\e}{4}\frac{(\vp\rq{})^2}{W}\Big|_{r_+}\right]\frac{{\cal A}}{4}\right)-\frac{{\cal A}}{4}T\delta\left(\frac{\e}{4}\frac{(\vp\rq{})^2}{W}\Big|_{r_+}\right).
\end{equation} 
The entropy is under the variation in the first term of the right hand side of the latter relation and takes the form \cite{Feng_PRD15}:
\begin{equation}\label{entropy}
S=\left(1+\frac{\e}{4}\frac{(\vp\rq{})^2}{W}\Big|_{r_+}\right)\frac{{\cal A}}{4},
\end{equation}
whereas the remaining part of the right hand side of the relation (\ref{def_ent}) can be represented as follows:
\begin{equation}\label{Ph_dQ}
-\frac{{\cal A}}{4}T\delta\left(\frac{\e}{4}\frac{(\vp\rq{})^2}{W}\Big|_{r_+}\right)=\Phi^{+}_{\vp}\delta  Q^{+}_{\vp}
\end{equation}
and here $Q^{+}_{\vp}$ is the scalar \lq\lq{}charge\rq\rq{} at the horizon and $\Phi^{+}_{\vp}$ is the potential canonically conjugate the \lq\lq{}charge\rq\rq{} at the horizon. It should be pointed out that the written above relation does not lead to a unique expressions for the \lq\lq{}charge\rq\rq{} $Q^+_{\vp}$ as well as of the potential $\Phi^+_{\vp}$, they can be chosen with some arbitrariness. We take them in the form:
\begin{equation}\label{sc_pot}
Q^{+}_{\vp}=\omega_{n-1}\sqrt{1+\frac{\e}{4}\frac{(\vp\rq{})^2}{W}\Big|_{r_+}}, \quad \Phi^{+}_{\vp}=-\frac{{\cal A}T}{2\omega_{n-1}}\sqrt{1+\frac{\e}{4}\frac{(\vp\rq{})^2}{W}\Big|_{r_+}}
\end{equation}
The given expressions for the \lq\lq{}charge\rq\rq{} and potential differ from the corresponding relations taken in \cite{Feng_PRD15} but the both variants lead to the same left hand side of the relation (\ref{Ph_dQ}). The convenience of our variant will be shown in the following. It can be also verified that:
\begin{equation}\label{rel_entr_pot}
\Phi^{+}_{\vp}Q^{+}_{\vp}=-\frac{{\cal A}T}{2}\left(1+\frac{\e}{4}\frac{(\vp\rq{})^2}{W}\Big|_{r_+}\right)=-2TS
\end{equation}
Finally, the relation (\ref{TD_diff}) can be represented in the form:
\begin{equation}
\delta {\cal H}_{+}=T\delta S+\Phi^{+}_{\vp}\delta Q^{+}_{\vp}.
\end{equation} 
Due to the fact that the variations of the Hamiltonian ${\cal H}$ at the horizon and at the infinity should be equal to one another \cite{Wald_PRD93} we can write the first law in the form:
\begin{equation}\label{first_law}
\delta M=T\delta S+\Phi^{+}_{\vp}\delta Q^{+}_{\vp}.
\end{equation}
\subsection{Extended thermodynamics}
In this section we consider thermodynamic functions of the black hole using the so called extended phase space. This extension is based on assumption about the relation between the cosmological constant and thermodynamic pressure. For long time the cosmological constant $\L$ has been supposed to be held fixed. Approximately a decade ago it was suggested that the cosmological constant might be varied and could be treated on the equal footing with the parameters that had been undergone the variation like mass, entropy, charge and angular momentum in case the black hole was charged or rotating. This idea relies on the suggestion that the cosmological constant is not a fundamental one, but it is rather caused by some field and possibly has quantum origin. The careful consideration showed that the quantity that appears as a conjugate to the variation of the cosmological constant would be proportional to some volume which was called as thermodynamic volume of the black hole, so the cosmological constant with suitably chosen constants can be identified with the  thermodynamic pressure \cite{Kastor_CQG09}:
\begin{equation}\label{pressure}
P=-\frac{\Lambda}{8\pi}
\end{equation}
Introduced thermodynamic pressure (\ref{pressure}) gives rise to the change of thermodynamic meaning of the black hole\rq{}s mass which should be identified not with the internal energy but rather with enthalpy. The conjugate value to the pressure, namely thermodynamic volume takes the form:
\begin{equation}
V=\left(\frac{\partial M}{\partial P}\right)_{S,Q^+_{\varphi}}.
\end{equation}
Having used the written above relation we write evident form  for the themodynamic volume. For the case of the odd dimension of space (odd $n$) it takes the form:
\begin{eqnarray}
\nonumber V=\frac{(n-1)\omega_{n-1}}{2}\left(\frac{(\al/\e+\L)}{2\al^2/\e^2}r^{n-2}_{+}+\frac{(\al/\e-\L)}{n(n-1)\al/\e}r^{n}_{+}-\frac{(\al/\e+\L)}{(n-1)\al/\e}\times\right.\\\left.\left[(-1)^{(n+1)/2}d^n\arctan\left(\frac{r_+}{d}\right)+\sum^{(n-1)/2}_{j=2}(-1)^jd^{2j}\frac{r^{n-2j}_{+}}{n-2j}\right]\right),
\end{eqnarray} 
whereas for the even $n$ we arrive at:
\begin{eqnarray}
\nonumber V=\frac{(n-1)\omega_{n-1}}{2}\left(\frac{(\al/\e+\L)}{2\al^2/\e^2}r^{n-2}_{+}+\frac{(\al/\e-\L)}{n(n-1)\al/\e}r^{n}_{+}-\frac{(\al/\e+\L)}{(n-1)\al/\e}\times\right.\\\left.\left[(-1)^{n/2}d^n\ln\left(\frac{r^2_+}{d^2}+1\right)+\sum^{(n-1)/2}_{j=2}(-1)^jd^{2j}\frac{r^{n-2j}_{+}}{n-2j}\right]\right).
\end{eqnarray} 
It is easy to notice that the only difference in these two expressions is due to inverse trigonometric and logarithmic terms which are inherited from the metric functions. 

In the case of the nonminimally coupled theory that is considered an additional intensive variable should be introduced \cite{Miao_EPJC16}:
\begin{equation}
\Pi=\frac{\al}{8\pi\e}.
\end{equation}
Corresponding canonically conjugate extensive value takes the form:
\begin{equation}
\Psi=\left(\frac{\partial M}{\partial\Pi}\right)_{S,Q^{+}_{\vp},P}
\end{equation} 
Having used the latter relation in case of odd $n$ we arrive at the expression:
\begin{eqnarray}
\nonumber\Psi=\frac{(n-1)\omega_{n-1}}{2}\left(\frac{\L(\al/\e+\L)}{2\al^3/\e^3}r^{n-2}_{+}+\frac{(\al^2/\e^2-\L^2)}{2n(n-1)\al^2/\e^2}r^n_{+}+\frac{(\al^2/\e^2-\L^2)}{2(n-1)\al^2/\e^2}\times\right.\\\nonumber\left[(-1)^{(n+1)/2}d^n\arctan\left(\frac{r_+}{d}\right)+\sum^{(n-1)/2}_{j=2}(-1)^jd^{2j}\frac{r^{n-2j}_{+}}{n-2j}\right]+\frac{(\al/\e+\L)^2}{2(n-1)\al^2/\e^2}\times\\\left.\left[(-1)^{(n+1)/2}d^n\left(-\frac{n}{2}\arctan\left(\frac{r_+}{d}\right)+\frac{r_{+}d}{2(r^2_{+}+d^2)}\right)-\sum^{(n-1)/2}_{j=2}(-1)^{j}jd^{2j}\frac{r^{n-2j}_{+}}{n-2j}\right]\right).
\end{eqnarray} 
For even $n$ we obtain:
 \begin{eqnarray}
\nonumber\Psi=\frac{(n-1)\omega_{n-1}}{2}\left(\frac{\L(\al/\e+\L)}{2\al^3/\e^3}r^{n-2}_{+}+\frac{(\al^2/\e^2-\L^2)}{2n(n-1)\al^2/\e^2}r^n_{+}+\frac{(\al^2/\e^2-\L^2)}{2(n-1)\al^2/\e^2}\times\right.\\\nonumber\left[(-1)^{n/2}\frac{d^n}{2}\ln\left(\frac{r^2_+}{d^2}+1\right)+\sum^{n/2-1}_{j=2}(-1)^jd^{2j}\frac{r^{n-2j}_{+}}{n-2j}\right]+\frac{(\al/\e+\L)^2}{2(n-1)\al^2/\e^2}\times\\\left.\left[(-)^{n/2}\frac{d^n}{2}\left(-\frac{n}{2}\ln\left(\frac{r^2_+}{d^2}+1\right)+\frac{r^2_{+}}{r^2_{+}+d^2}\right)-\sum^{n/2-1}_{j=2}(-1)^{j}jd^{2j}\frac{r^{n-2j}_{+}}{n-2j}\right]\right).
\end{eqnarray} 
Similarly as for the thermodynamic volume the obtained expressions have the only difference in logarithmic and arctangent functions. It appears to be strange that  the new thermodynamic variable, namely the variable $\Pi$ has been introduced, but it should be noted that for example in case of charged black hole in Einstein-Born-Infeld theory to develop consistent extended thermodynamics it was assumed that the Born-Infeld coupling constant  should be varied in complete analogy with the cosmological constant \cite{Gunasekaran_JHEP12}.

The introduced thermodynamic values allow us to write the extended first law in the following form:
\begin{equation}\label{first_law_gen}
\delta M=T\delta S+\Phi^{+}_{\vp}\delta Q^{+}_{\vp}+V\delta P+\Psi\delta\Pi.
\end{equation}
Taking into consideration the relation (\ref{rel_entr_pot}) we can write the generalized Smarr relation:
\begin{equation}\label{Smarr}
 (n-2)M=(n-1)TS-2VP-2\Psi\Pi.
\end{equation}
It is worth been emphasized that the obtained Smarr relation has its grounds in two facts, namely the chosen form of the \lq\lq{}scalar\rq\rq{} potential (\ref{sc_pot}) and the idea of extended thermodynamic phase space which gives possibility to introduce new thermodynamic variables such as $P$ and $\Pi$ and the values conjugate to them. We also note that the form of the scalar \lq\lq{}potential\rq\rq{} chosen in the paper \cite{Feng_PRD15} satisfies the first law (\ref{first_law}) but does not allow to obtain the Smarr relation even in case of the extended phase space. The Smarr relation (\ref{Smarr}) was also obtained in the work \cite{Miao_EPJC16} but there the entropy was assumed to take the same form as in the standard General Relativity and it does not match with the Wald approach.  
\subsection{Heat capacity and Gibbs free energy} 
\begin{figure}
\centerline{\includegraphics[scale=0.33,clip]{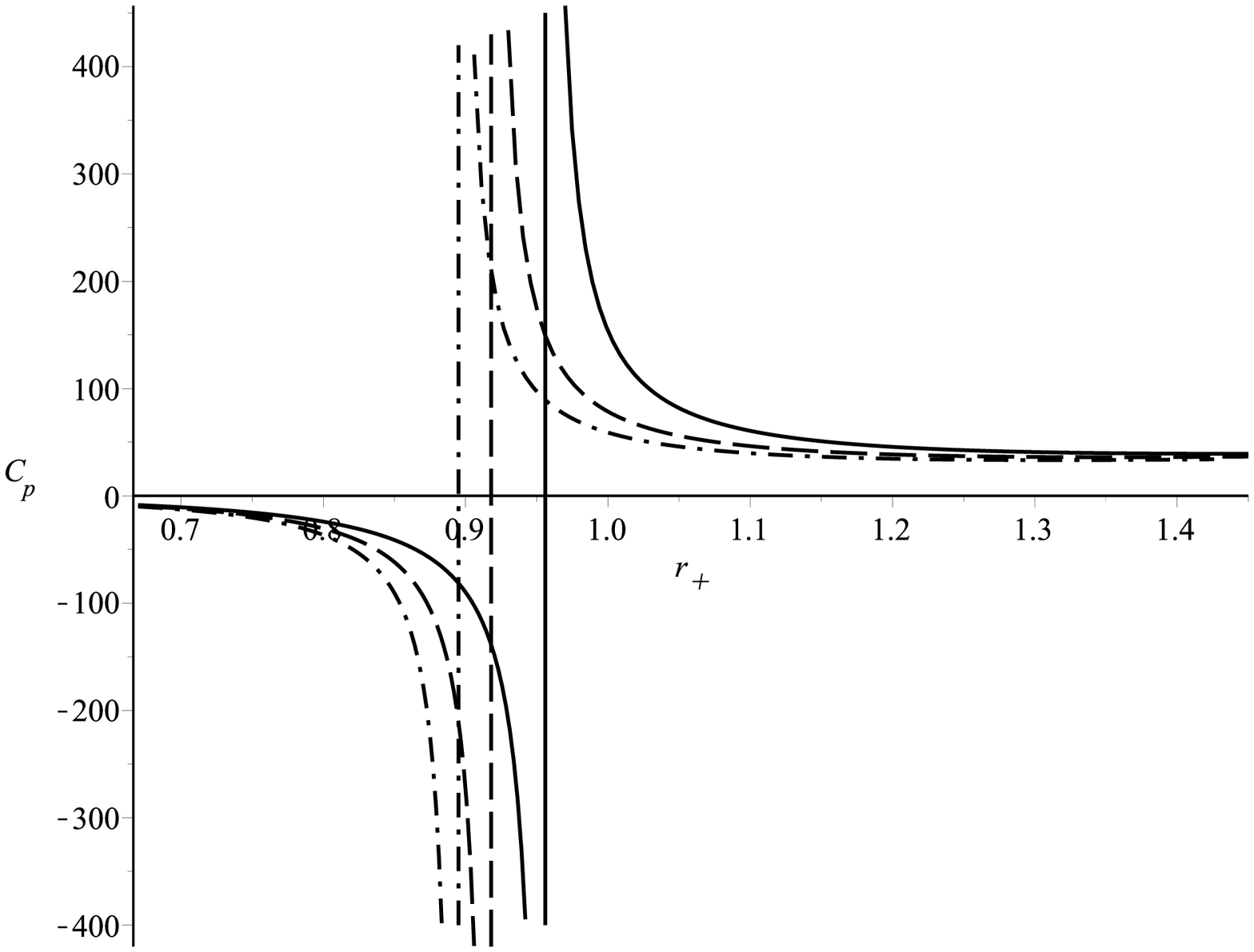}\includegraphics[scale=0.33,clip]{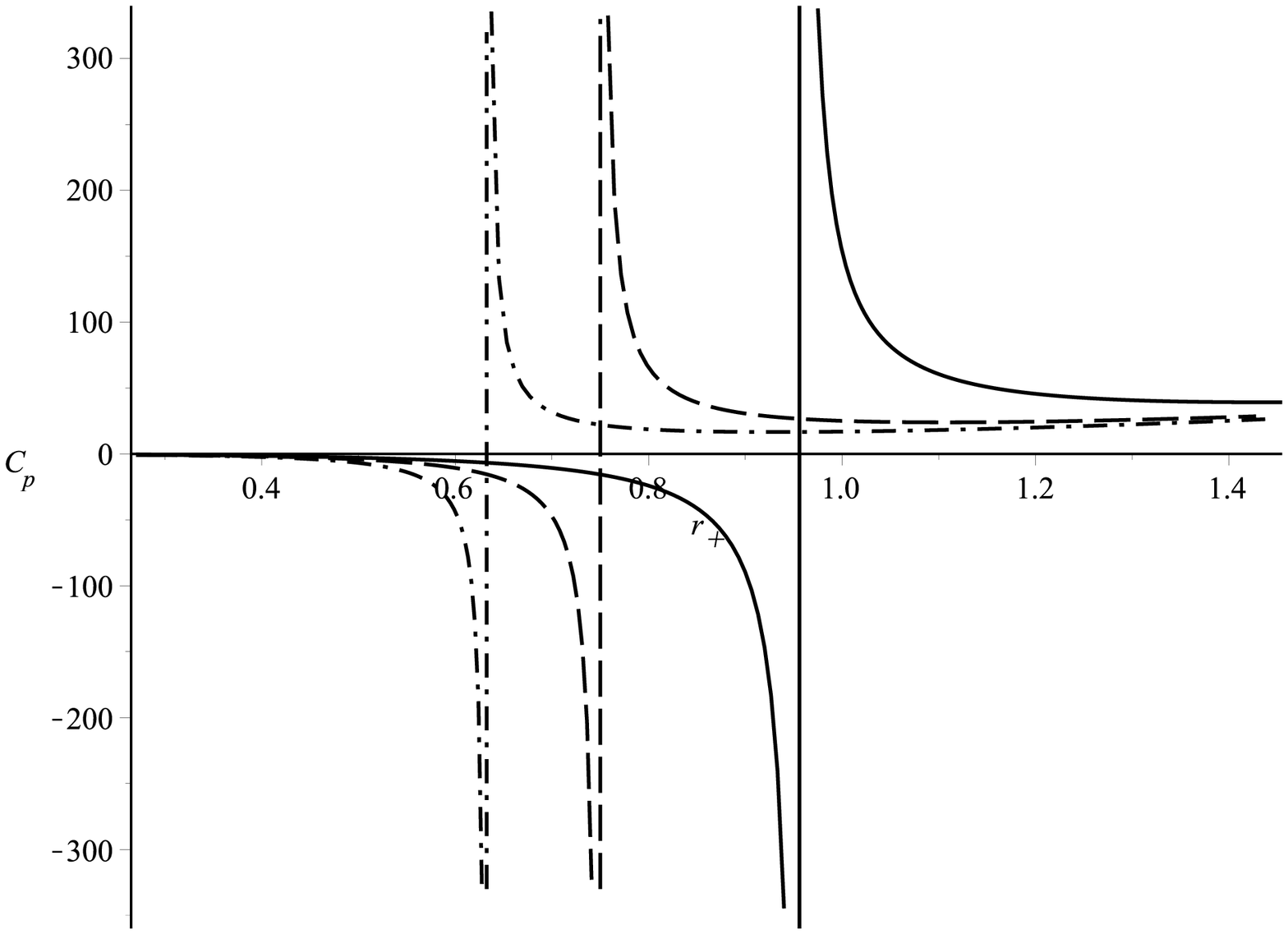}}
\caption{Heat capacity $C_p$ as a function of $r_+$ for several values of the parameters $\e$ (the left graph) and $\L$ (the right one). For the left graph the solid, dashed and dash-dotted curves correspond to $\e=0.4$, $\e=0.6$ and $\e=0.8$ respectively and the other parameters are equal to: $n=3$, $\al=0.2$, $\L=-1$. For the right graph the solid, dashed and dash-dotted curves correspond to $\L=-1$, $\L=-1.5$ and $\L=-2$ respectively and $n=3$, $\al=0.2$, $\e=0.4$}\label{Cp_diff_e}
\end{figure}

To understand thermodynamic behaviour better one can analyze heat capacity and Gibbs free energy. The heat capacity can be calculated as follows:
\begin{equation}\label{C_P}
C_{P}=T\left(\frac{\partial S}{\partial T}\right)_{P,\Pi,Q^{+}_{\vp}}=T\left(\frac{\partial S}{\partial r_+}\right)_{P,\Pi,Q^+_{\vp}}\left(\frac{\partial r_+}{\partial T}\right)_{P,\Pi,Q^+_{\vp}}
\end{equation}
It should be noted that here other parameters such as $\Pi$ and $Q^{+}_{\vp}$ are also held fixed. As a result the heat capacity can be written in the form:
\begin{eqnarray}
\nonumber C_P=\frac{(n-1)\omega_{n-1}}{4}\frac{((\al-\L\e)r^2_{+}+\e(n-1)(n-2))}{(2\al r^2_{+}+\e(n-1)(n-2))}\left(\frac{(\al-\L\e)^2}{2(n-1)\al\e}r^{n-1}_{+}+\left(1-\frac{(\al+\L\e)^2}{4\al^2}\right)\times\right.\\\nonumber\left.(n-2)r^{n-3}_{+}+\frac{(\al+\L\e)^2}{2(n-1)\al\e}\left[(-1)^{\frac{(n+1)}{2}}\frac{d^{n+1}}{(r^2_{+}+d^2)}+\sum^{(n-1)/2}_{j=2}(-1)^jd^{2j}r^{n-1-2j}_{+}\right]\right)\times\\\nonumber
\left(\frac{(\al-\L\e)^2}{2(n-1)\al\e}-\left(1-\frac{(\al+\L\e)^2}{4\al^2}\right)\frac{(n-2)}{r^2_+}+\frac{(\al+\L\e)^2}{2(n-1)\al\e}\times\right.\\\left.\left[(-1)^{(n+1)\over{2}}\frac{d^{n+1}r^{1-n}_{+}((2-n)d^2-nr^2_{+})}{(r^2_{+}+d^2)^2}+\sum^{(n-1)/2}_{j=2}(-1)^jd^{2j}(1-2j)r^{-2j}_{+}\right]\right)^{-1}
\end{eqnarray} 
for odd $n$ and
\begin{eqnarray}
\nonumber C_P=\frac{(n-1)\omega_{n-1}}{4}\frac{((\al-\L\e)r^2_{+}+\e(n-1)(n-2))}{(2\al r^2_{+}+\e(n-1)(n-2))}\left(\frac{(\al-\L\e)^2}{2(n-1)\al\e}r^{n-1}_{+}+\left(1-\frac{(\al+\L\e)^2}{4\al^2}\right)\times\right.\\\nonumber\left.(n-2)r^{n-3}_{+}+\frac{(\al+\L\e)^2}{2(n-1)\al\e}\left[(-1)^{\frac{n}{2}}\frac{d^nr_+}{(r^2_{+}+d^2)}+\sum^{n/2-1}_{j=2}(-1)^jd^{2j}r^{n-1-2j}_{+}\right]\right)\times\\\nonumber
\left(\frac{(\al-\L\e)^2}{2(n-1)\al\e}-\left(1-\frac{(\al+\L\e)^2}{4\al^2}\right)\frac{(n-2)}{r^2_+}+\frac{(\al+\L\e)^2}{2(n-1)\al\e}\times\right.\\\left.\left[(-1)^{n\over{2}}\frac{d^{n}r^{2-n}_{+}((3-n)d^2+(1-n)r^2_{+})}{(r^2_{+}+d^2)^2}+\sum^{n/2-1}_{j=2}(-1)^jd^{2j}(1-2j)r^{-2j}_{+}\right]\right)^{-1}
\end{eqnarray}
for even $n$.
The dependence of the heat capacity $C_p$ from the horizon radius $r_+$  is represented on the figures (\ref{Cp_diff_e}). The figures show the discontinuity of heat capacities typical for Hawking-Page phase transitions and as it has been mentioned before it takes place due to the non monotonous behaviour of temperature. The left graph of Fig.(\ref{Cp_diff_e}) demonstrates that with increase of parameter $\e$ we have the decrease of the horizon radius $r_+$ for which discontinuity happens and it means that the greater $\e$ is the black holes with smaller radius of horizon $r_+$ would be thermodynamically stable. The same situation takes place with the increase of the absolute  value of the cosmological constant $\L$ what is demonstrated on the right graph of the Fig.(\ref{Cp_diff_e}). It should also be noted that for the horizon radius $r_+$ large enough  in comparison with the discontinuity radius the heat capacity $C_p$ monotonically increase with the following rise of $r_+$.

The Gibbs free energy is defined as follows:
 \begin{equation}\label{gibbs_pot}
 G=M-TS
 \end{equation} 
 Having used this standard relation we can write:
 \begin{eqnarray}
\nonumber  G=\frac{\omega_{n-1}}{16\pi}\left(\left(1-\frac{(\al+\L\e)^2}{4\al^2}\right)r^{n-2}_{+}-\frac{(\al-\L\e)^2}{2n(n-1)\al\e}r^n_{+}+\frac{(\al+\L\e)^2}{2(n-1\al\e)}\right.\\\left.\left[(-1)^{(n+1)\over{2}}d^n\left((n-1)
 \arctan\left(r_+\over{d}\right)-\frac{r_{+}d}{r^2_{+}+d^2}\right)+\sum^{(n-1)/2}_{j=2}(-1)^jd^{2j}\frac{(2j-1)}{n-2j}r^{n-2j}_{+}\right]\right)
 \end{eqnarray}
 for odd $n$ and 
  \begin{eqnarray}
\nonumber  G=\frac{\omega_{n-1}}{16\pi}\left(\left(1-\frac{(\al+\L\e)^2}{4\al^2}\right)r^{n-2}_{+}-\frac{(\al-\L\e)^2}{2n(n-1)\al\e}r^n_{+}+\frac{(\al+\L\e)^2}{2(n-1\al\e)}\right.\\\left.\left[(-1)^{n\over{2}}d^n\left(\frac{(n-1)}{2}
 \ln\left(\frac{r^2_+}{d^2}+1\right)-\frac{r^2_+}{r^2_{+}+d^2}\right)+\sum^{n/2-1}_{j=2}(-1)^jd^{2j}\frac{(2j-1)}{n-2j}r^{n-2j}_{+}\right]\right)
 \end{eqnarray}
 for even $n$.
 
\begin{figure}
\centerline{\includegraphics[scale=0.33,clip]{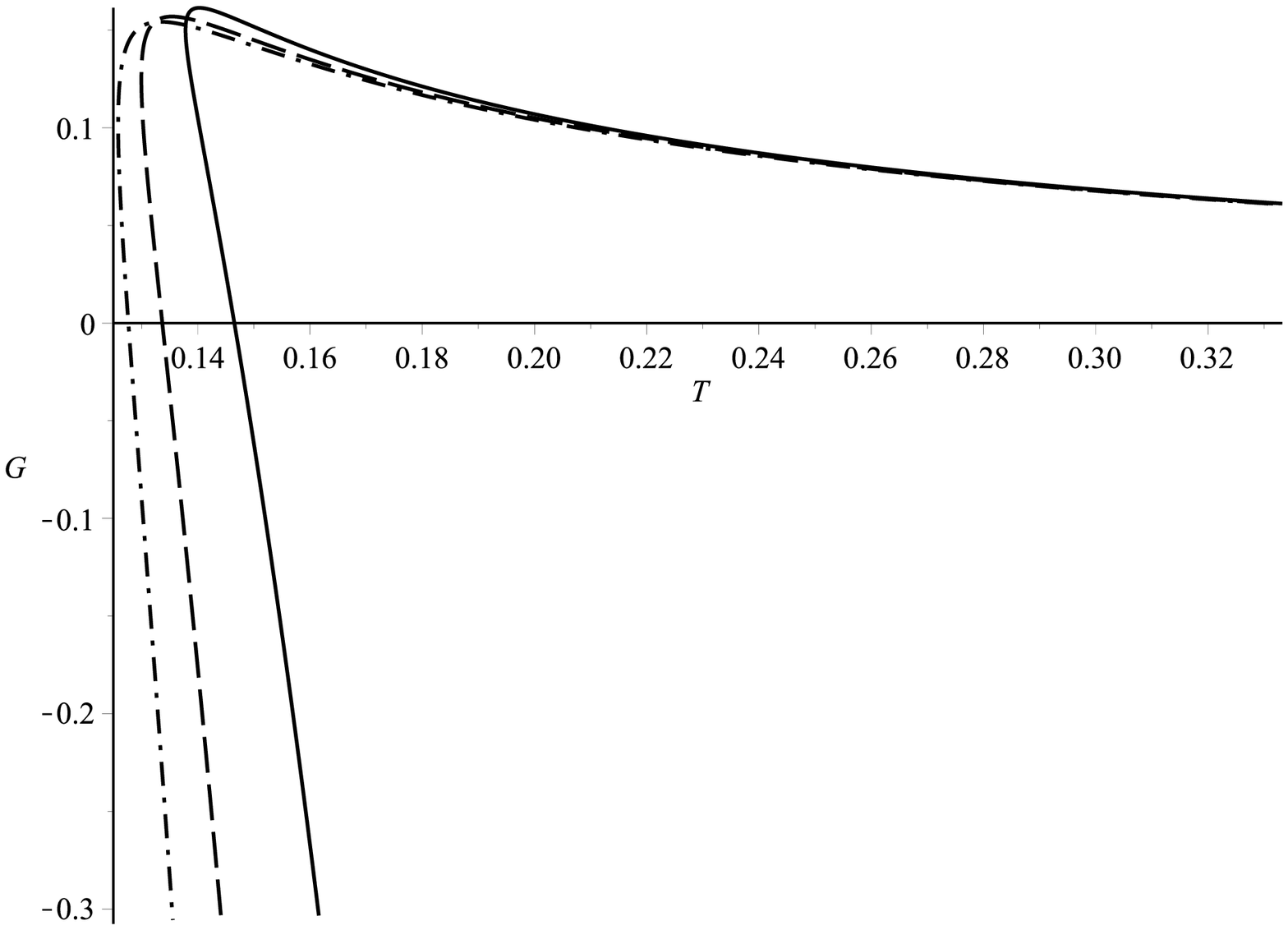}\includegraphics[scale=0.33,clip]{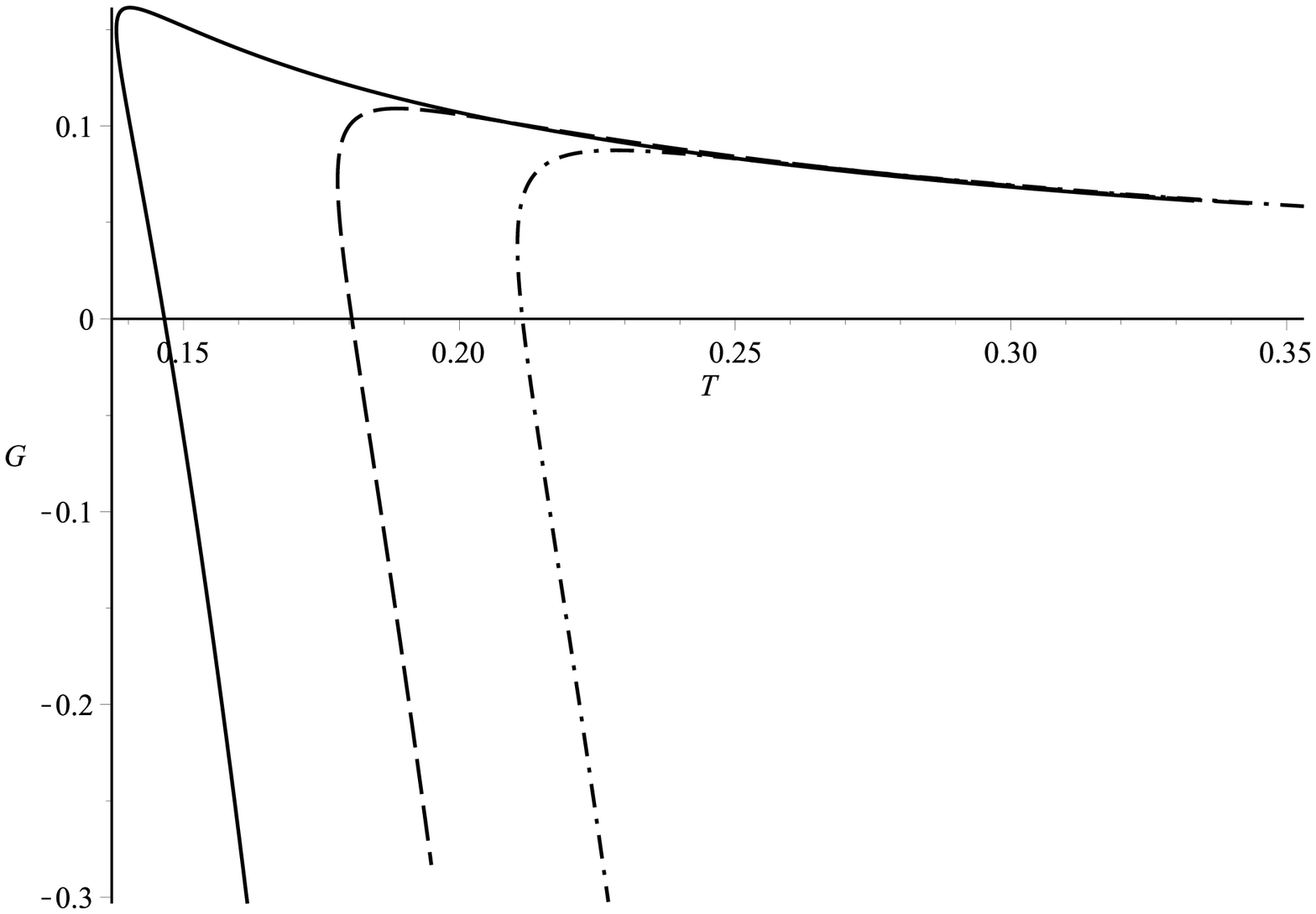}}
\caption{Gibbs free energy $G$ as a function of temperature $T$ for several values of the parameter $\e$ (the left graph) and $\L$ (the right one).  Namely for the left graph the solid, dashed and dash-dotted curves correspond to $\e=0.4$, $\e=0.6$ and $\e=0.8$ respectively and other parameters are fixed: $n=3$, $\al=0.2$, $\L=-1$. For the right graph  the solid, dashed and dash-dotted curves correspond to $\L=-1$, $\L=-1.5$ and $\L=-2$ respectively and other parameters are equal to: $n=3$, $\al=0.2$, $\e=0.4$.}\label{Gibbs_1}
\end{figure}

\begin{figure}
\centerline{\includegraphics[scale=0.33,clip]{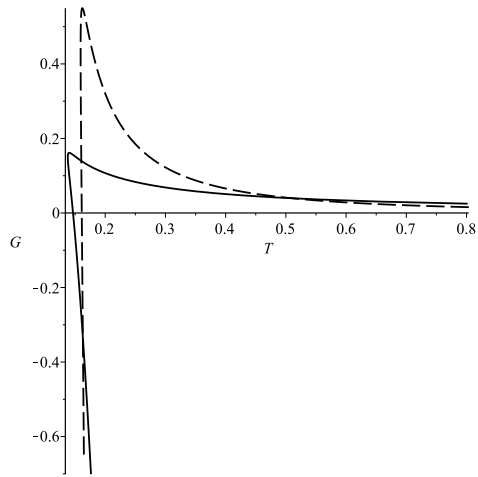}}
\caption{Gibbs free energy $G$ as a function of temperature $T$ for different dimensions  of space. The solid and dashed curves correspond to $n=3$ and $n=4$ respectively.  The other parameters are equal to: $\al=0.2$, $\e=0.4$ and $\L=-1$.}\label{Gibbs_2}
\end{figure}

The Gibbs free energy as a function of temperate is represented on the figures (\ref{Gibbs_1}). The Gibbs free energy has typical behaviour for a black hole with a Hawking-Page phase transition which can be explained by the fact that for large $r$ the metric has AdS-like behaviour which is necessary for the appearance of the Hawking-Page phase transition, the behaviour of the metric functions for small distances is of the less importance for it. It should be noted that similar behaviour of the Gibbs free energy was obtained in the paper \cite{Miao_EPJC16}. The Fig.(\ref{Gibbs_2}) shows that the behaviour of the Gibbs free energy as a function of temperature is qualitatively the same for different dimensions of the space $n$, but for larger $n$ the peak of the Gibbs free energy becomes higher whereas for small temperatures it has steeper descent in comparison with the case of smaller $n$. For large temperatures the Gibbs free energy decreases slowly and does not have substantial differences for different values of $n$.

 \section{Conclusions}
 The work is devoted to the investigation of a slowly rotating black black hole solution in the theory with nonminimal derivative coupling. We also take into account the cosmological constant. It should be pointed out that a static solution in this approach was also considered in the works \cite{Minamitsuji_PRD2014,Anabalon_PRD14} and our solution is in agreement with those previously obtained solutions. The careful analysis shows that we have two types of solutions depending on the sign of  the parameter $\eta$, whereas the other coupling parameter of the theory $\al$ is assumed to have positive value. 
 
 The first type of solution, namely when $\e>0$ is valid when the cosmological constant is negative ($\L<0$). The negative sign of the cosmological constant is necessary to provide positivity of kinetic energy term of the scalar field in the outer part of the black hole solution. In general, we can conclude, that this type of solution represented by the relations (\ref{funct_u_odd}) and (\ref{funct_u_even}) shows the behaviour similar to AdS-Schwarzschild black hole, namely it has the only event horizon and demonstrates the AdS-like behaviour for large distances, whereas for small distances it recovers the AdS-Schwarzschild part with additional terms, proportional to $\sim r^4$. It is worth being pointed out that in the regime of large nonminimal coupling (large $\e$) one can also arrive at the AdS-Schwarzschild part with additional term $\sim r^4$ but in this case this correction does not depend on the parameters $\al$ and $\e$ in contrast with the situation for small $r$, where some dependence takes place.  It should also be noted that the first type of solution is represented by the mentioned above two relations (\ref{funct_u_odd}) and (\ref{funct_u_even}) describing the cases of odd and even space dimensions respectively. The difference between odd and even $n$ lies in the appearance of the $\sim\arctan(r/d)$ term for the first case and $\sim\ln(r^2/d^2+1)$  for the second one. But, it should be stressed that the mentioned terms give rise to similar behaviour of the metric functions (\ref{funct_u_odd}) and (\ref{funct_u_even}) for small and large distances as well as in the regime of strong nonminimal coupling. The additional fact to confirm this statement is the behaviour of the Kretschmann scalar (\ref{Kr_scalar}) which for both parities of $n$ behaves in the same manner for small as well as for large distances.
 
 The second type of solution takes place for the negative parameter $\e$, namely here we arrive at the relations (\ref{funct_u_odd_neg}) and (\ref{f_u_neg_even}).  The careful analysis of this type of solution has shown that it cannot be treated as a black hole solution. The most important fact that demonstrates why we do not have a black hole is the existence of the divergence point $r_b$ for the kinetic term of the scalar field $\vp$ which lies between the roots of the metric function $U(r)$ and it would mean that this scalar field would be unstable somewhere in the outer domain of the black hole.
 
 We also considered thermodynamics of the obtained black hole. Firstly we obtained black hole's temperature using the standard procedure based on Killing vectors (\ref{temp_gen}) and it is supposed to be true even in more general cases than standard General Relativity. The obtained relations for the temperature, namely (\ref{T_odd}) and (\ref{T_even}) demonstrate that the temperature has nonmonotonous behaviour as a function of horizon radius. It grows up to infinity when the radius of horizon goes to zero. For large horizon radius it increases almost linearly which is typical for Schwarzschild-AdS black hole. For some intermediate value of $r_+$ the temperature achieves its minimal values which tells us that we have Hawking-Page phase transition that separates thermodynamically stable and unstable solutions.  To sum up the analysis of the temperature we can conclude that qualitatively its behaviour is very similar to the mentioned above Schwarzschild-AdS black hole.

It should be pointed out that in case of the theory with nonminimal derivative coupling or more general Horndeski theory there some ambiguities in the definition of important thermodynamic values such as entropy \cite{Feng_JHEP15,Feng_PRD15}. It was shown \cite{Feng_JHEP15,Feng_PRD15} that the most substantial approach to the definition of entropy is based on the method proposed by Wald \cite{Wald_PRD93,Iyer_PRD94} which relies upon the Noether conserved quantities. In our case we also used Wald procedure and obtained the expressions for the black hole's mass (\ref{bh_mass}), entropy (\ref{entropy}) and the first law (\ref{first_law}). The entropy we obtained is proportional to the area of horizon ${\cal A}$ but has some factor which appears due to nonminimal coupling and the obtained relation is in full agreement with corresponding formulas in \cite{Feng_JHEP15,Feng_PRD15}. To obtain the first law it was proposed to introduce additional scalar ``charge'' $Q^+_{\vp}$, related to the field $\vp$ but it can be chosen in different ways. The choice we have made here allows us to derive the Smarr relation (\ref{Smarr}) which is not possible to obtain, when one uses the form of the scalar potential, given in \cite{Feng_PRD15}. 

Having introduced thermodynamic pressure (\ref{pressure}) we have also examined the thermodynamics of the black hole in so called extended phase space. The extended phase space allowed us to construct  the Smarr relation (\ref{Smarr}). To obtain this relation additional intensive variable $\Pi$ was introduced \cite{Miao_EPJC16}. It should be noted that that appearance of additional intensive variables related to some coupling constants is typical when one considers extended thermodynamic phase space, for example in the case of Einstein-Born-Infeld theory \cite{Gunasekaran_JHEP12} where it was introduced an additional thermodynamic variable related to the Born-Infeld coupling constant. We also point out that the Smarr relation we have obtained does not include the scalar ``charge'' $Q^+_{\vp}$ because it is not a conserved quantity. At the same time its variation is present in the first law (\ref{first_law}) and its generalized form (\ref{first_law_gen}). Finally we have calculated heat capacity (\ref{C_P}) and Gibbs free energy (\ref{gibbs_pot}). The heat capacity has a discontinuity point which separate stable and unstable phases. The Gibbs free energy is nonmonotonous function of temperature and its behaviour is in general similar to what we have for Schwarzschild-AdS black hole. 
\section{Acknowledgements}
This work was partly supported by Project FF-30F (No. 0116U001539) from the Ministry of Education and Science of Ukraine.

\end{document}